# Properties and Radial Evolution of Solar Wind Turbulence Near Mercury's Orbit

Xinmin Li[1], Chuanfei Dong[1,2], Lina Z. Hadid[3], Sae Aizawa[3], Chi Zhang[1], Hongyang Zhou[1], Liang Wang[1], Jiawei Gao[1], and James A. Slavin[4]

1. *Center for Space Physics and Department of Astronomy, Boston University, Boston, MA 02215, USA*
2. *School of Natural Sciences, Institute for Advanced Study, Princeton, NJ 08540, USA*
3. *Laboratoire de Physique des Plasmas (LPP), CNRS, Observatoire de Paris, Sorbonne Université, Université Paris Saclay, Ecole polytechnique, Institut Polytechnique de Paris, Palaiseau, 91120, France*
4. *Department of Climate and Space Sciences and Engineering, University of Michigan, Ann Arbor, MI 48109, USA*

Corresponding author: X. Li (xli8@bu.edu) and C. Dong (dcfy@bu.edu)

**Abstract**

We present a comprehensive statistical study of the radial evolution of solar wind turbulence near Mercury's orbit using long-term magnetic field measurements from the MESSENGER mission. Owing to Mercury's highly elliptical orbit and the spacecraft's repeated, extended residence in the upstream solar wind, the data set provides more than 17,000 hours of observations, enabling robust statistics across well-defined heliocentric distance intervals (0.31–0.47 au). We find that inertial-range spectral slopes remain close to −3/2 throughout Mercury's orbit, showing no significant radial evolution. Combined with low magnetic compressibility, this result indicates a stable, predominantly Alfvénic inertial-range cascade already established here. In contrast, kinetic-range spectral slopes exhibit clear radial evolution, becoming progressively shallower with increasing heliocentric distance, highlighting the greater sensitivity of kinetic-scale turbulence to heliocentric conditions. The ion-scale spectral break frequency decreases with distance in the spacecraft frame, while its normalized form increases relative to the local proton cyclotron frequency, demonstrating that the break is not tied to proton cyclotron frequency but reflects evolving local plasma conditions. Magnetic compressibility shows a similar frequency dependence at all distances, with a subtle radial enhancement of compressive fluctuations at kinetic scales. Autocorrelation analysis reveals strong anisotropy, with the correlation times of field-aligned magnetic fluctuations increasing with heliocentric distance, while those of perpendicular fluctuations remain shorter and nearly invariant. Together, these results demonstrate a clear scale-dependent radial evolution of solar wind turbulence near Mercury's orbit, providing new constraints on the development of kinetic processes in the inner heliosphere.

## 1. Introduction

The solar wind is a continuously expanding plasma outflow from the Sun that carries magnetic field fluctuations across a wide range of spatial and temporal scales throughout solar system. Through nonlinear interactions, these fluctuations develop into turbulence characterized by an energy cascade from large magnetohydrodynamic (MHD) scales to ion and electron kinetic scales. This cascade establishes an inertial range governed by nonlinear energy transfer (Boldyrev 2005; Dong et al. 2018; Dong et al. 2022; Goldstein, Roberts, & Matthaeus 1995; Schekochihin 2022), followed by a transition near ion characteristic scales that commonly manifests as a spectral break in the magnetic field power spectrum (Alexandrova et al. 2012; Alexandrova et al. 2009; Chen et al. 2010). Beyond this break, turbulence enters the kinetic range, where kinetic effects modify fluctuation properties through changes in cascade mechanisms and enhanced dissipation (Bruno & Carbone 2013; Tu & Marsch 1995). Long-term observations, at near-Earth distances around 1 astronomical unit (au), have established a well-defined reference picture of solar wind turbulence. Magnetic field power spectra typically exhibit an inertial-range slope close to −5/3 (Chen et al. 2013; Chen et al. 2011; Goldstein, et al. 1995; Smith et al. 2006), followed by a steeper kinetic-range slope, commonly ranging between −2.5 and −4.0, with a peak occurrence around −2.8 (Alexandrova, et al. 2012; Alexandrova, et al. 2009; Kiyani, Osman, & Chapman 2015; Sahraoui et al. 2013). The spectral break between inertial and kinetic ranges is often observed at frequencies of order 0.1-1.0 Hz, and has been linked to multiple physical mechanisms operating near ion kinetic scales (Chen et al. 2014; Goldstein et al. 2015; Terres & Li 2022; Vech et al. 2018; Woodham et al. 2018). Beyond the spectrum, turbulence at 1 au is characterized by low magnetic compressibility in the inertial range, enhanced compressibility toward kinetic scales, and pronounced scale-dependent anisotropy relative to the local mean magnetic field (Dasso et al. 2005;

Kiyani et al. 2013; Matthaeus, Goldstein, & Roberts 1990; Osman & Horbury 2007; Wang, Tu, & He 2019; Wicks et al. 2012; Zhao et al. 2022).

Above properties only represent conditions at a single heliocentric distance. During the radial expansion of the solar wind, both the plasma and the embedded magnetic field evolve systematically; consequently, turbulence properties are also expected to vary systematically with heliocentric distance (Bruno & Carbone 2013). Numerous studies have investigated the radial evolution of solar wind turbulence using observations from multiple spacecraft at different heliocentric distances (Alberti et al. 2022a; Alberti et al. 2022b; Duan et al. 2020; He et al. 2013; Lotz et al. 2023; Perri, Carbone, & Veltri 2010; Shi et al. 2021; Zhao, et al. 2022; Zhu et al. 2025). However, the reported quantitative trends are not yet fully converged, as many results are derived from short-duration intervals, limited orbital coverage, or different analysis approaches across different missions, which can introduce systematic inconsistencies and hinder direct quantitative comparisons. Addressing these limitations requires long-term, continuous observations within a well-defined heliocentric distance range and a uniform analysis framework.

The MErcury Surface, Space ENvironment, GEochemistry, and Ranging (MESSENGER) mission provides a unique opportunity to study solar wind turbulence in the inner heliosphere. As illustrated in Figure 1(e), Mercury's orbit spans heliocentric distances from approximately 0.31 to 0.47 au, corresponding to nearly 50% radial variation within a short orbital period of about 88 days. Owing to Mercury's small magnetosphere (Dong et al. 2019; Slavin et al. 2008; Winslow et al. 2013), MESSENGER spent roughly half of each orbit in the upstream solar wind (an example is shown in Figures 1(a)-1(d) and 1(f)), enabling repeated and long-term sampling across this radial range. This repeated sampling within a well-defined heliocentric distance interval enables statistically well-converged analyses of turbulence properties. While spanning a more confined radial range than multi-mission surveys, the MESSENGER observations allow a systematic investigation of turbulence evolution near Mercury's orbit. Here, we perform a statistical analysis of solar wind turbulence near Mercury's orbit using long-term MESSENGER magnetic field measurements (~17,400 hours of solar wind data), focusing on the radial evolution of spectral slopes, spectral break frequencies, magnetic compressibility, and correlation scales derived from large, well-converged data sets. Our results provide quantitative characterization of how these turbulence properties evolve in this key inner-heliospheric region (0.31 to 0.47 au) and improve our understanding of the turbulent plasma conditions that directly impact Mercury's space environment.

## 2. Data and Methods

### 2.1 Data and Instruments

This study is based on magnetic field measurements from the MESSENGER spacecraft over its entire orbital mission at Mercury. We use vector magnetic field data acquired by the onboard magnetometer (MAG), which provides continuous measurements in both Mercury's magnetosphere and the upstream solar wind, with a time resolution of 20 Hz (Anderson et al. 2007).

Solar wind intervals are identified using an independent database of bow shock and magnetopause crossings compiled for the MESSENGER mission (Sun 2026). This database enables a reliable separation of upstream solar wind intervals from Mercury's magnetosphere. To conservatively select undisturbed solar wind conditions, magnetic field data within 10 minutes of each bow shock crossing are excluded. Applying these criteria to the full MESSENGER mission yields approximately 17,400 hours of upstream solar wind data, providing a statistically robust data set for the analysis.

## 2.2 Analysis Methods

For the analysis, the solar wind magnetic field data are divided into consecutive 30-min intervals. Within each interval, we examine the properties of magnetic field fluctuations, including power spectral characteristics, magnetic compressibility, and autocorrelation properties.

The power spectral density (PSD) of magnetic field fluctuations is calculated using the Welch method (Welch 1967). For each 30-min interval, the magnetic field time series is divided into 10-min segments with 50% overlap. A Hanning window is applied to each segment to reduce spectral leakage, and the background magnetic field is removed prior to the Fourier transform. The single-sided PSD is obtained by averaging the periodograms of all segments. Unless otherwise specified, the PSD discussed hereafter refers to the trace of the spectral matrix constructed from the three components of the magnetic field fluctuations. With these parameters, the analyzed frequency range spans approximately 2 mHz to 10 Hz. Given that the typical proton cyclotron frequency is in the order of 0.1 Hz (Huang et al. 2020), this frequency range adequately covers both the inertial and kinetic ranges, allowing reliable identification of the spectral break.

To identify spectral breaks and determine spectral slopes, we apply a fitting procedure following the method introduced in our previous work (Li et al. 2025). Briefly, the PSD and frequency (from 0.01 Hz to 5 Hz) are transformed into log–log space and fitted using either single or double power law models. The lower limit of 0.01 Hz is chosen to minimize contamination from energy ejection range, which is typically observed below $10^{-3}$ Hz near Mercury's orbit (Pradata et al. 2025). As this study focuses on fully developed turbulence, only intervals whose spectra are well described by a double–power-law distribution are retained. Intervals in which the PSD deviates significantly from power-law behavior, for example due to coherent structures such as discontinuities, current sheets, or narrowband wave activity, are excluded from the subsequent statistical analysis, following approaches commonly adopted in previous studies of solar wind turbulence (Huang, et al. 2020; Lotz, et al. 2023).

Magnetic compressibility is quantified by $C_{\parallel}(f)$ (Gary & Smith 2009), defined as

$$C_{\parallel}(f) = \frac{|\delta B_{\parallel}(f)|^2}{|\delta B_{\parallel}(f)|^2 + |\delta B_{\perp 1}(f)|^2 + |\delta B_{\perp 2}(f)|^2} \tag{1}$$

where $\delta B_{\parallel}$ denotes the magnetic field fluctuation parallel to the background magnetic field, and $\delta B_{\perp 1}$and $\delta B_{\perp 2}$ denote the two perpendicular components.

The autocorrelation function (ACF) is calculated as

$$A_{\mathrm{norm},L} = \frac{N}{N-L} \frac{\sum_{n=1}^{N-L} X_n X_{n+L}}{\sum_{n=1}^{N} X_n^2} \tag{2}$$

where $X_n$ represents $\delta B_{\parallel}$, $\delta B_{\perp 1}$, or $\delta B_{\perp 2}$, and $L$ is the time lag (Bruno & Carbone 2013). The correlation time is defined as the time lag at which the ACF first decreases to 1/*e*, where *e* is the base of the natural logarithm.

Both $C_{\parallel}$ and the ACF are evaluated in a local field aligned coordinate system, in which the background magnetic field is obtained using a 1-minute sliding average.

As an example, Figures 1(g)-1(i) shows the PSD, $C_{\parallel}(f)$, and ACF for the interval 12:25:30-12:55:30 UT, highlighted by the gray shading in Figures 1(a)-1(d). The PSD exhibits a clear spectral break at around 0.13 Hz, close to the local proton cyclotron frequency (0.15 Hz). This break separates the inertial and kinetic ranges, with fitted power-law slopes of −1.60 and −2.64, respectively. Magnetic compressibility (Figure 1(h)) shows a strong frequency dependent. At low frequencies, $C_{\parallel}(f)$ is much smaller than 1/3, indicating that the fluctuations are predominantly transverse and weakly compressive. Toward higher frequencies, $C_{\parallel}(f)$ increases but remains below 1/3, suggesting an enhanced contribution of compressive fluctuations at kinetic scales, although transverse fluctuations remain dominant. Figure 1(i) shows the ACF of the magnetic field fluctuations. The parallel component exhibits a significantly longer correlation time than the perpendicular components, indicating a pronounced anisotropy of the turbulence relative to the background magnetic field. This example illustrates the key spectral, compressibility, and correlation properties extracted from an individual 30-min interval; in the following sections, we statistically investigate the radial evolution of these quantities with heliocentric distance.

## 3 Results

Figure 2 shows the radial dependence of magnetic field spectral slopes and spectral break frequencies of solar wind turbulence near Mercury's orbit. Figure 2(a) shows the distributions of spectral slopes in the inertial range as functions of heliocentric distance from 0.31 to 0.47 au. Two horizontal dashed lines indicate the −3/2 spectrum (Boldyrev 2006; Iroshnikov 1963; Kraichnan 1965) and the −5/3 Kolmogorov spectrum (Kolmogoroff 1941). The spectral slopes are narrowly distributed around a median value of approximately −1.5 across all radial bins. And the median inertial-range slope shows no systematic radial dependence within the statistical uncertainties. In contrast, the kinetic-range (Figure 2(b)) spectral slopes exhibit a broader distribution and a clear radial trend. The median kinetic-range slope becomes progressively shallower with increasing heliocentric distance, evolving from values near −3.3 near perihelion to about −3.0 near aphelion. The dashed curve represents a power-law fit of the form $ar^b$ to the median kinetic-range spectral slopes, with fitted parameters a=−2.4 and b=−0.27, indicating a gradual radial flattening of the kinetic-range spectrum.

Figures 2(c) and 2(d) show the radial dependence of the spectral break frequency. In Figure 2(c), the break frequency decreases with increasing heliocentric distance. The dashed curve indicates a power-law fit to the median break frequencies ($ar^b$, where a=0.6 and b=−0.37), highlighting the decreasing radial trend. When normalized by the local proton cyclotron frequency $f_{cp}$ (Figure 2(d)), an opposite trend is observed. The normalized break frequency increases with heliocentric distance, rising from values of approximately 2.0 $f_{cp}$ near perihelion to about 3.5 $f_{cp}$ near aphelion. The dashed curve in Figure 2(d) likewise represents a power-law fit $ar^b$, with a=9.0 and b=1.28.

Figure 3 presents the statistical distributions of magnetic compressibility $C_{\parallel}$ as a function of frequency. The data are divided into six radial groups according to heliocentric distance. For each radial group, the color-coded map represents the distribution of $C_{\parallel}$ at a given frequency, constructed from all solar wind intervals within that radial bin. The color intensity indicates the occurrence frequency of $C_{\parallel}$ , with darker colors corresponding to a larger number of events clustered at that value. Across all radial groups, $C_{\parallel}$ shows a broadly similar behavior. At low frequencies, the distributions are concentrated at $C_{\parallel} \ll 1/3$, indicating predominantly transverse and weakly compressive fluctuations in the inertial range. With increasing frequency, the distributions shift toward larger $C_{\parallel}$ values but still less than 1/3, reflecting an enhanced contribution of compressive fluctuations.

The radial dependence of $C_{\parallel}$ is weak but systematic, with $C_{\parallel}$ gradually increasing with heliocentric distance. To better illustrate this trend, we examine the distributions at two representative frequencies, 0.5 Hz and 5.0 Hz (Figures 3(g) and 3(h)), where different colors denote different heliocentric distance ranges. The $C_{\parallel}$ distributions show an overall shift toward larger values with increasing heliocentric distance. These trends indicate that, although transverse fluctuations remain dominant throughout Mercury's orbital range, compressive fluctuations become increasingly important with radial distance, particularly at kinetic scales

Figure 4 shows the radial dependence of the ACFs of magnetic field fluctuations near Mercury's orbit. Figures 4(a)–4(f) display the ACFs averaged over all 30-min intervals within each radial group, calculated separately for the parallel and perpendicular components relative to the local mean magnetic field. The horizontal dashed line marks the 1/e level used to define the correlation time. Across all radial groups, the ACFs exhibit a pronounced anisotropy. The parallel component decays substantially more slowly than the perpendicular components, indicating significantly longer correlation times along the background magnetic field direction. The two perpendicular components are nearly indistinguishable and largely overlap with each other in all radial bins, indicating that the turbulence is statistically axisymmetric in the plane perpendicular to the background magnetic field. This anisotropic and axisymmetric correlation structure is consistently observed throughout Mercury's orbital range.

Figure 4(g) shows the radial dependence of the correlation times derived from the ACFs. The correlation times of the perpendicular components are shorter than those of the parallel component and remain nearly constant with heliocentric distance. In contrast, the correlation time of the parallel component is significantly larger and increases with heliocentric distance. These results indicate that the radial evolution of temporal correlation properties is confined primarily to the field-aligned component, while the perpendicular correlation times remain essentially invariant over Mercury's orbital range.

## 4 Discussions and Conclusion

Our statistical results show that, throughout Mercury's orbit (∼0.30–0.47 au), the inertial-range magnetic-field spectra exhibit no significant radial evolution, with spectral slopes remaining close to −3/2. Recent studies based on Parker Solar Probe observations, spanning heliocentric distances of roughly 0.1–1.0 au, have investigated the radial evolution of inertial-range spectral slopes in near-Sun solar-wind turbulence. While some analyses report that the inertial-range slope remains nearly invariant within the inner heliosphere (Lotz, et al. 2023), others find a systematic steepening of the slope with increasing heliocentric distance, although the reported variations within the 0.1–1 au range do not appear to be strictly monotonic (Chen et al. 2020; Zhao, et al. 2022). Over the more limited heliocentric range examined here, but based on a large statistical sample, our results show that within Mercury's orbit (~0.37–0.47 au) the inertial-range slope exhibits no significant radial evolution.

Solar-wind turbulence near 1 AU is more commonly observed to exhibit a steeper inertial-range spectrum close to -5/3. The -5/3 slope is often associated with Kolmogorov-like turbulence, in which strong nonlinear interactions drive an efficient turbulent energy cascade across scales. In comparison, the shallower -3/2 spectrum has been interpreted in terms of Alfvénic MHD turbulence, including scale-dependent dynamic alignment models (Boldyrev 2006). In these models, velocity and magnetic fluctuations become progressively aligned toward smaller scales, which weakens the nonlinear interaction and leads to a shallower turbulent cascade. The persistence of a -3/2 spectrum throughout Mercury's orbit may therefore suggest that inertial range of solar-wind turbulence in the inner heliosphere remains in a more Alfvénic and dynamically aligned state closer to the Sun.

Kinetic-scale spectra exhibit clear radial evolution near Mercury's orbit, with spectral slopes becoming progressively shallower with increasing heliocentric distance. This behavior, consistent with previous near-Sun (0.1-0.7 au) observations (Lotz, et al. 2023), indicates that the kinetic-scale cascade is more sensitive to heliocentric conditions than inertial-range turbulence. One contributing factor may be the radial decrease in turbulence intensity: closer to the Sun, enhanced turbulence supports stronger nonlinear interactions and more efficient transfer to small scales, producing steeper kinetic-range spectra, whereas the gradual weakening of turbulence with increasing heliocentric distance results in spectral flattening (Smith, et al. 2006; Zank et al. 2021). A second factor is the temporal evolution, or "age," of the turbulence. As the solar wind expands, kinetic-scale fluctuations have more time to redistribute and relax, favoring shallower slopes (Lotz, et al. 2023).

Previous studies have reported different conclusions regarding the radial evolution of the ion-scale spectral break frequency. Analyses focused on the inner heliosphere (~0.3–0.7 au) generally found that the break frequency decreases with increasing heliocentric distance (Bruno & Trenchi 2014; Duan, et al. 2020; Lotz, et al. 2023), whereas studies spanning a much broader radial range (up to 4.9 au) reported little or no systematic radial trend (Bourouaine et al. 2012; Perri, et al. 2010). Using long-term MESSENGER observations, our results demonstrate that, at least within Mercury's orbital range, the spectral break frequency decreases with heliocentric distance. This trend is consistent with the radial evolution of characteristic ion kinetic scales as the solar wind expands, leading to the transition from magnetohydrodynamic to kinetic regimes occurring at progressively lower frequencies in the spacecraft frame.

When normalized by the local proton cyclotron frequency $f_{cp}$, the ratio $f_b/f_{cp}$ increases with heliocentric distance across Mercury's orbit. This behavior indicates that the spectral break is not fixed at a constant fraction of $f_{cp}$. Previous studies have suggested that different dissipation mechanisms may place the spectral break near different characteristic ion scales or frequencies, such as $f_{cp}$, proton inertial length, or proton gyroradius (Bourouaine, et al. 2012; Perri, et al. 2010). However, due to the lack of continuous plasma measurements from MESSENGER, we can only conclude that the observed break frequency does not exhibit a simple proportional relationship with $f_{cp}$.

Magnetic compressibility exhibits broadly similar behavior throughout Mercury's orbital range. In all radial bins, inertial-range fluctuations are characterized by low compressibility ($C_{\parallel} \ll 1/3$), indicating predominantly transverse and weakly compressive, Alfvénic turbulence. This property is consistent with previous observations on solar wind across a wide range of heliocentric distances (Bruno & Carbone 2013; Tu & Marsch 1995) as well as with observations in planetary magnetosheaths (Hadid et al. 2015; Hadid et al. 2018; Huang, et al. 2020). At kinetic scales, magnetic compressibility increases with frequency, reflecting an enhanced contribution of compressive fluctuations, in agreement with earlier studies linking kinetic-range turbulence to kinetic Alfven waves (KAWs) and other compressive modes (e.g. oblique whistler-mode fluctuations) (Chen, et al. 2010; Salem et al. 2012). Moreover, we also observe a subtle but systematic radial enhancement in the high-$C_{\parallel}$ tail, especially in the kinetic range, indicating a gradual increase in the relative contribution of compressive fluctuations as the solar wind expands.

Correlation times in the solar wind generally increase with heliocentric distance and exhibit clear anisotropy, with longer correlations along the local mean magnetic field than in the perpendicular components (Chen, et al. 2020; Ruiz et al. 2014; Tu & Marsch 1995), and our results near Mercury's orbit are consistent with this general picture. Across all radial bins, the correlation times of perpendicular magnetic fluctuations are much shorter than those of the parallel component, indicating a strong anisotropy of turbulent structures. In addition, we find that the field-aligned correlation time exhibits a clear radial increase across Mercury's orbital range, whereas the perpendicular correlation times remain nearly invariant. Given that the solar wind speed does not vary systematically over this distance range (McGregor et al. 2011; Venzmer & Bothmer 2018), the observed correlation times can be qualitatively related to spatial correlation scales. These results suggest that radial expansion primarily affects field-aligned, compressive structures, while the transverse, largely incompressible component of the turbulence has already reached a statistically stable state near Mercury's orbit.

It should be noted that, because MESSENGER does not provide solar wind velocity measurements, the observed radial trends may be affected by Doppler-shift effects associated with variations in the solar wind speed. In particular, the spectral break frequencies analyzed in this study are measured in the spacecraft frame and may therefore be influenced by changes in the solar wind speed. However, previous studies, including Parker Solar Probe observations, have shown that the solar wind speed does not exhibit a significant monotonic variation over the heliocentric distance range considered here (McGregor, et al. 2011; Venzmer & Bothmer 2018), implying that the observed increase of $f_b/f_{cp}$ with heliocentric distance is unlikely to be solely caused by systematic radial variations in solar wind speed. In addition, previous studies have shown that different solar wind streams may exhibit different turbulence properties (Bruno & Carbone 2013). However, the relatively stable solar wind speed over the heliocentric distance range considered here suggests that the dominant solar wind populations do not vary systematically with distance. In addition, the large statistical dataset used in this study reduces the influence of individual extreme events or short-term solar wind variations. Therefore, the observed radial trends are unlikely to simply result from systematically different solar wind populations. Finally, although the influence of small-scale coherent structures has been largely removed, the effects of large-scale structures with timescales much longer than the 10 min analysis window, such as coronal mass ejections (CMEs), may still remain to some extent.

In summary, using long-term magnetic field measurements from MESSENGER, we investigate the properties and radial evolution of solar wind turbulence around Mercury orbit (0.31-0.47 au). We find that the inertial-range spectral slopes remain close to −3/2 with no significant radial evolution near Mercury's orbit, whereas kinetic-range slopes become progressively shallower with increasing heliocentric distance. Magnetic compressibility stays low in the inertial range but increases at kinetic scales, with a weak radial enhancement. Correlation analysis further reveals increasing field-aligned fluctuations' correlation times with distance, while perpendicular correlations remain short and nearly invariant, indicating a clear scale-dependent radial evolution of near-Sun solar wind turbulence.

**Acknowledgments**

This work is supported by NASA grant 80NSSC23K0894, DOE grant DE-SC0024639, the Alfred P. Sloan Research Fellowship, and the IBM Einstein Fellow Fund at the Institute for Advanced Study, Princeton. We thank the entire MESSENGER team and instrument principal investigators for providing and calibrating data, and the calibrated MESSENGER MAG magnetic field data in MSO coordinates were obtained from the NASA Planetary Data System (Korth 2021). Mercury's position and spacecraft ephemerides are calculated using the SPICE toolkit (https://naif.jpl.nasa.gov/naif/toolkit.html)

## Figures and Figure Captions

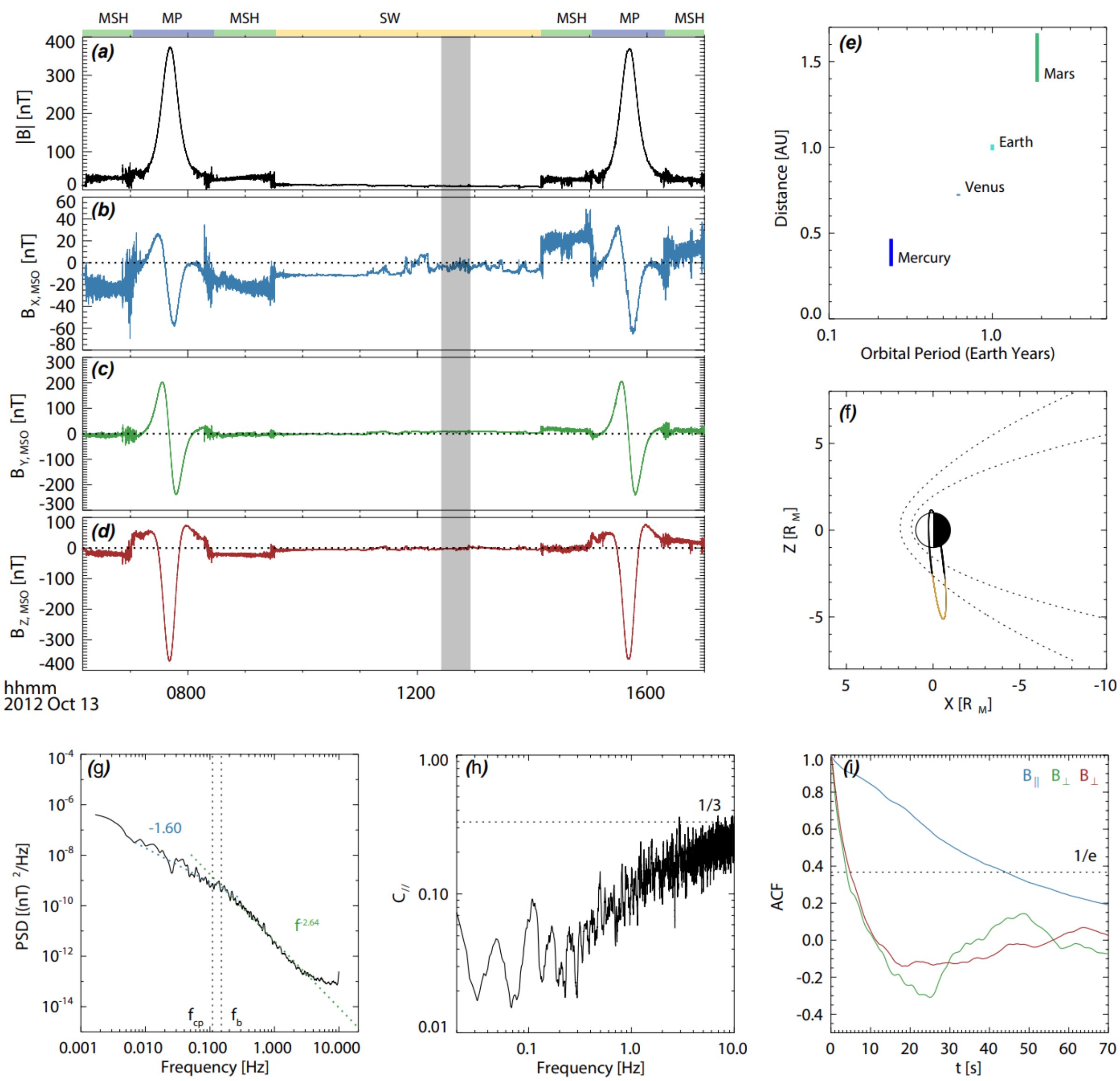

**Figure 1. Overview of a complete MESSENGER orbit around Mercury.** Panels (a)–(d) show the magnetic field magnitude |B| and the three magnetic field components in the Mercury Solar Orbital (MSO) coordinate system on 13 October 2012. Colored bars at the top indicate different plasma regions, with green denoting the magnetosheath, blue the magnetosphere, and yellow the solar wind. Panel (e) shows the heliocentric distances and orbital periods of Mercury, Venus, Earth,

and Mars. Panel (f) shows the trajectory of MESSENGER relative to Mercury for the interval in panels (a)–(d), with dashed curves marking the bow shock and the magnetopause. Panels (g)–(i) show the power spectral density (PSD), magnetic compressibility $C_{\parallel}$, and the autocorrelation functions (ACFs) of magnetic field fluctuations calculated for 12:25:30–12:55:30 UT (highlighted by the gray shading). In panel (g), the PSD exhibits a spectral break and fitted slopes of −1.60 and −2.64 in the inertial and kinetic ranges. Vertical dashed lines indicate the average proton cyclotron frequency and the spectral break frequency. In panel (i), the blue curve denotes the parallel component, while the red and green curves denote the perpendicular components.

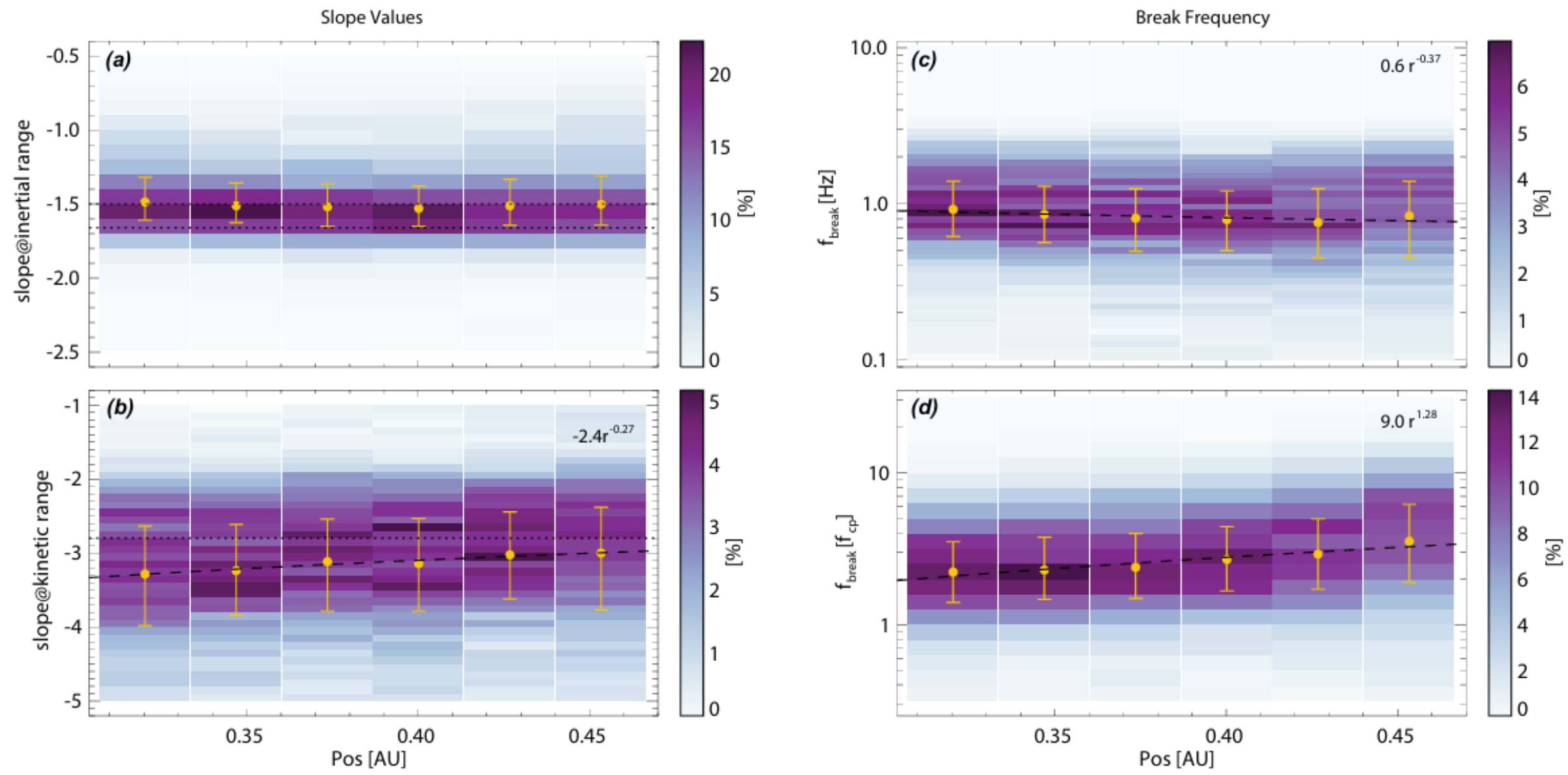

**Figure 2. Radial dependence of spectral slopes and the spectral break frequency of solar wind turbulence near Mercury's orbit.** (a) and (b) show the distributions of magnetic field spectral slopes in the inertial and kinetic ranges, respectively, as functions of heliocentric distance from 0.31 to 0.47 AU. Dots denote the median values, and error bars represent the first and third quartiles. (c) and (d) show the radial dependence of the spectral break frequency. In (c), the break frequency is shown in Hz, while in (d) it is normalized by the proton cyclotron frequency $f_{cp}$. Dashed curves indicate power law fits of the form $ar^b$.

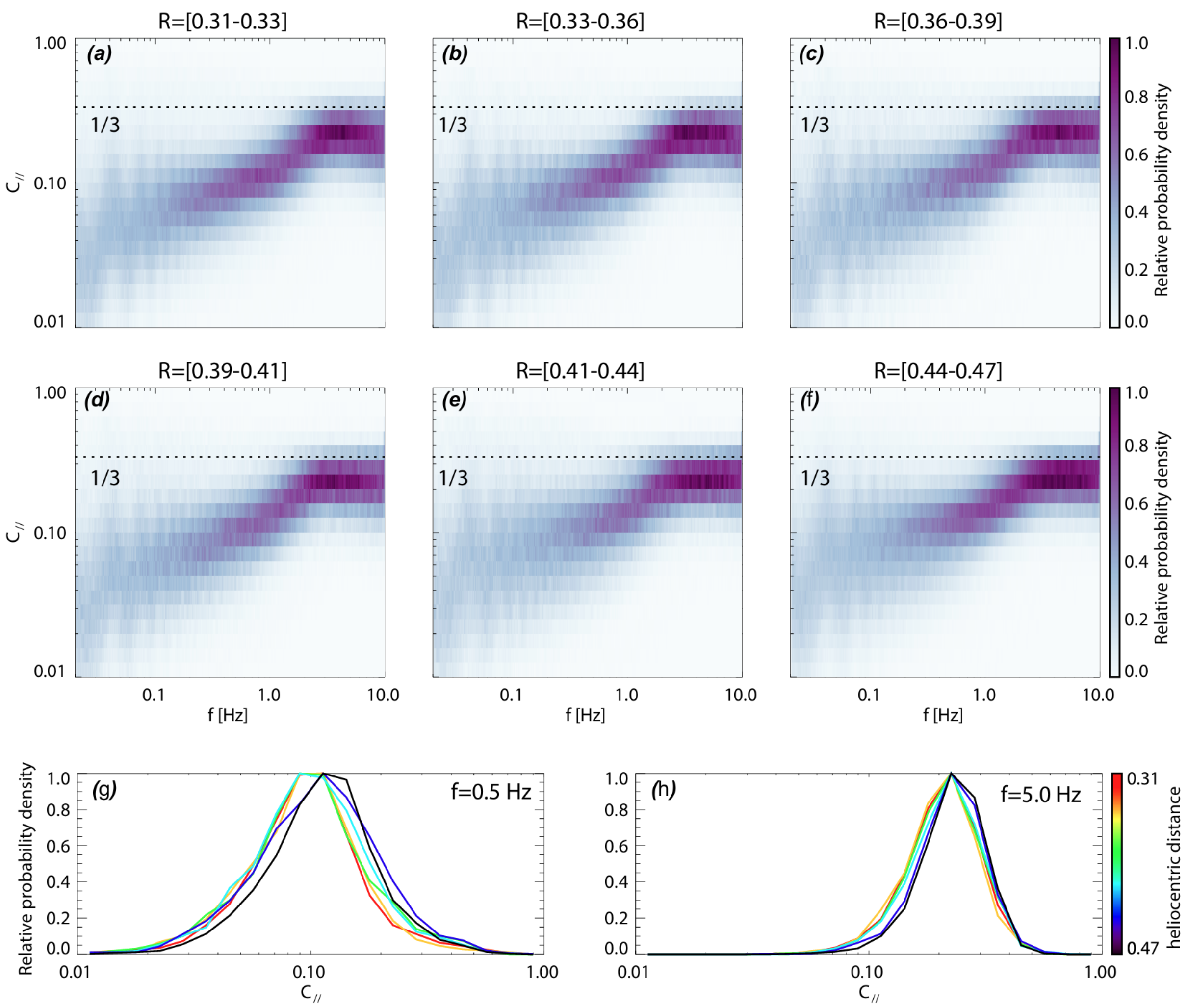

**Figure 3. Radial dependence of magnetic compressibility of magnetic field fluctuations near Mercury's orbit.** (a)–(f) show the distributions of the magnetic compressibility $C_{\parallel}$ as functions of frequency for six heliocentric distance ranges from 0.31 to 0.47 AU. The horizontal dashed line at $C_{\parallel}$ =1/3 indicates isotropic fluctuations. (g) and (h) present the distributions of $C_{\parallel}$ at fixed frequencies of 0.5 Hz and 5.0 Hz, respectively, for different heliocentric distances.

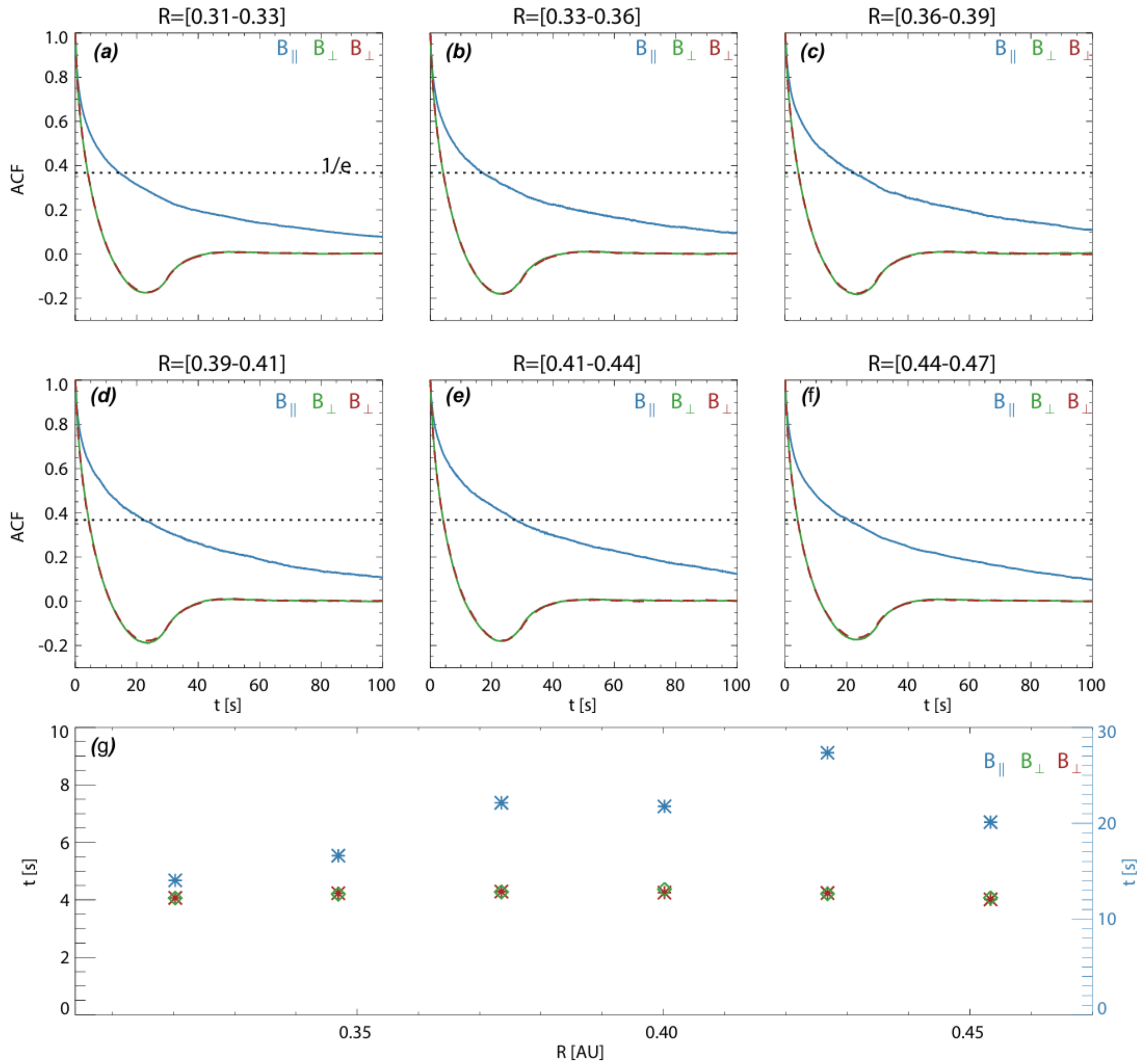

**Figure 4. Radial dependence of autocorrelation properties and characteristic timescales of magnetic field fluctuations near Mercury's orbit.** (a)–(f) show the autocorrelation functions (ACFs) of magnetic field fluctuations for six heliocentric distance ranges from 0.31 to 0.47 au. The horizontal dashed line indicates the 1/e level used to define the correlation time. Blue curves denote fluctuations parallel to the background magnetic field, while red and green curves denote the two perpendicular components. (g) summarizes the correlation times as functions of heliocentric distance. The perpendicular components (red asterisks and green diamonds) are

plotted against the left axis, whereas the parallel component (blue asterisks), which has a much larger correlation time, is plotted against the right axis.